\documentclass[12pt, a4paper]{article}
\usepackage{graphicx}
\newcommand{\bea}{\begin{eqnarray}}
\newcommand{\eea}{\end{eqnarray}}
\usepackage{amsmath} 
\usepackage{cite}
\input{epsf} 
\newcommand\as{\alpha_{\mathrm{S}}}

\def\beq{\begin{equation}} 
\def\eeq{\end{equation}} 
\def\to{\rightarrow}

\def\b0{\beta_0}

\def\beeq{\begin{eqnarray}}
\def\eeeq{\end{eqnarray}}

\def\ca{{\cal A}}
\def\sh{\hat{s}}
\def\Re{{\rm Re}}
\def\Im{{\rm Im}}

\begin{document} 

\begin{titlepage}
\renewcommand{\thefootnote}{\fnsymbol{footnote}}
\begin{flushright}
\end{flushright}
\par \vspace{10mm}

\begin{center}
{ \Large \bf A complete ${\cal O}(\as^2)$ calculation of the signal-background \\[0.5cm]
              interference for the Higgs diphoton decay channel}
\end{center}
\par \vspace{2mm}
\begin{center}
{\bf Daniel de Florian}\footnote{deflo@df.uba.ar},
{\bf Nerina Fidanza}\footnote{nfidanza@df.uba.ar},
{\bf Roger Hern\'andez-Pinto}\footnote{roger@df.uba.ar},
{\bf Javier Mazzitelli}\footnote{jmazzi@df.uba.ar},
{\bf Yamila Rotstein Habarnau}\footnote{yrotstein@df.uba.ar} and
{\bf Germ\'an Sborlini}\footnote{gfsborlini@df.uba.ar}\\

\vspace{5mm}

Departamento de F\'isica, FCEyN, Universidad de Buenos Aires, \\
(1428) Pabell\'on 1, Ciudad Universitaria, Capital Federal, Argentina\\

\vspace{5mm}

\end{center}

\par \vspace{2mm}
\begin{center} {\large \bf Abstract} \end{center}
\begin{quote}
\pretolerance 10000

We present the full ${\cal O}(\as^2)$ computation of the interference effects between the Higgs diphoton signal and the continuum background at the LHC.  While the main contribution to the interference originates on the $gg$ partonic subprocess, we find that the corrections from the $qg$ and $q\bar{q}$ channels amount up to 35\% of it.  
We discuss the effect of these new subprocesses in the shift of the diphoton invariant mass peak recently reported by S. Martin in Ref.\cite{Martin:2012xc}.

\end{quote}

\vspace*{\fill}
\begin{flushleft}
March 2013
\end{flushleft}
\end{titlepage}

\setcounter{footnote}{1}
\renewcommand{\thefootnote}{\fnsymbol{footnote}}


Recent discoveries of a new particle presented both by CMS \cite{:2012gu} and ATLAS \cite{:2012gk} collaborations at the Large Hadron Collider (LHC) are consistent with the Standard Model Higgs boson. Up to now, its mass is estimated to be around $125.3\pm 0.4(\text{stat})\pm 0.5(\text{syst})\,\text{GeV}$ by CMS and $126.0\pm 0.4(\text{stat})\pm 0.4(\text{syst})\,\text{GeV}$ by ATLAS, though the experimental uncertainty could be reduced to $100\,\text{MeV}$ \cite{ATLASTDR} after collecting enough luminosity. 

The resonance observed in the reconstruction of the diphoton invariant mass in proton-proton collisions at the LHC turns out to be one of the main discovery channels and, therefore, requests precise theoretical calculations for the corresponding cross section. 

The dominant mechanism for SM Higgs boson production at hadron colliders is gluon-gluon fusion \cite{Georgi:1977gs}, through a heavy-quark (mainly top-quark) loop. The QCD radiative corrections to the total cross section have been computed at the next-to-leading order (NLO) in Refs.~\cite{Dawson:1990zj,Djouadi:1991tk,Spira:1995rr}, and at the next-to-next-to-leading order (NNLO) accuracy in \cite{Harlander:2002wh,Anastasiou:2002yz,Ravindran:2003um}. NNLO results at the exclusive level can be found in Refs.~\cite{Anastasiou:2005qj,Anastasiou:2007mz,Catani:2007vq,Grazzini:2008tf}.
State of the art computations for this channel \cite{deFlorian:2012yg} include electroweak corrections at NLO \cite{ew,Actis:2008ug} and soft gluon resummation to next-to-next-to leading logarithmic accuracy \cite{Catani:2003zt}\footnote{For a review see \cite{Dittmaier:2011ti,Dittmaier:2012vm}.}. 

The rare decay $H\rightarrow \gamma\gamma$ is also mediated by ($W$ and heavy quark) loops \cite{Ellis:1975ap,Shifman:1979eb}. Corrections to the corresponding branching ratio are known at NLO accuracy for both QCD
\cite{Spira:1995rr,Zheng:1990qa,Djouadi:1990aj,Dawson:1992cy,Djouadi:1993ji,Melnikov:1993tj,Inoue:1994jq} and electroweak \cite{Actis:2008ug} cases. Missing higher orders are estimated to be below 1\%.

The corresponding background for diphoton production has also been recently computed up to NNLO \cite{Catani:2011qz}, but the interference between {\it signal} and {\it background} has not yet been evaluated to such level of accuracy.

The interference of the resonant process $ij \to X+H ( \to \gamma \gamma )$ with the continuum QCD background $ij \to X+\gamma\gamma $ induced by quark loops can be expressed at the level of the partonic cross section as:
\begin{eqnarray}
\delta\hat{\sigma}_{ij\to X+ H\to \gamma\gamma} &=& 
-2 (\sh-m_H^2) { \Re \left( \ca_{ij\to X+H} \ca_{H\to\gamma\gamma} 
                          \ca_{\rm cont}^* \right) 
        \over (\sh - m_H^2)^2 + m_H^2 \Gamma_H^2 }
\nonumber\\
&& 
-2 m_H \Gamma_H { \Im \left( \ca_{ij\to X+H} \ca_{H\to\gamma\gamma} 
                          \ca_{\rm cont}^* \right)
        \over (\sh - m_H^2)^2 + m_H^2 \Gamma_H^2 } \,,
\label{intpartonic}
\end{eqnarray}
where $\sh$ is the partonic invariant mass, $m_H$ and $\Gamma_H$ are the Higgs mass and decay width respectively \footnote{The details on the implementation of the lineshape \cite{Goria:2011wa} have a very small effect on the light Higgs discussed in this note. We rely here on a naive Breit-Wigner prescription.}.

As pointed out in \cite{Dixon:2003yb,Dicus:1987fk}, given that the contribution arising from the real part of the amplitudes is odd in $\sh$ around $m_H$, its effect on the total $\gamma\gamma$ rate is subdominant. For the gluon-gluon partonic subprocess, Dicus and Willenbrock \cite{Dicus:1987fk} found that the imaginary part of the corresponding one-loop amplitude has a quark mass suppression for the relevant helicity combinations. 
Dixon and Siu \cite{Dixon:2003yb} computed the main contribution of the interference to the cross-section, which originates on the two-loop imaginary part of the continuum amplitude $gg \to \gamma\gamma$. 
Recently, Martin \cite{Martin:2012xc} showed that even though the real part hardly contributes at the cross-section level, it has a quantifiable effect on the position of the diphoton invariant mass peak, producing a shift of $ {\cal O}(100\,\text{MeV})$ towards a lower mass region, once the smearing effect of the detector is taken into account.

The $gg$ interference channel considered in \cite{Martin:2012xc} is not the only ${\cal O}(\as^2)$ contribution that has to be taken into account for a full understanding of the interference term, since other partonic subprocesses initiated by $qg$ and $q\bar{q}$ can contribute at the same order. While these channels are suppressed with respect to the $gg$ subprocess for the Higgs signal, they dominate the $\gamma\gamma$ QCD background, and therefore their contribution to the interference can not be $\textit{a priori}$ neglected. At variance with the $gg$ subprocess that necessarily requests at least a one-loop amplitude for the background, the contribution from the remaining channels arises from tree-level amplitudes, and can therefore only contribute to the real part of the interference in Eq.(\ref{intpartonic}) \footnote{Apart from a small imaginary part originated on the heavy-quark loop in the Higgs production amplitude that is discarded in this note since we rely on the effective $ggH$ vertex. There is also an imaginary contribution in the decay process $H \to \gamma \gamma$ which was included since the full expression for the vertex was used, but was found to be negligible.}.

In this note we present the results obtained for the remaining $qg$ and $q\bar q$ channels, finalizing a full (lowest order) ${\cal O}(\as^2)$ calculation of the interference between Higgs diphoton decay amplitude and the corresponding continuum background. We concentrate on the effect of the new interference channels on the position of the diphoton invariant mass peak.

\begin{figure} 
\begin {center}
\qquad \qquad \includegraphics[scale=0.55]{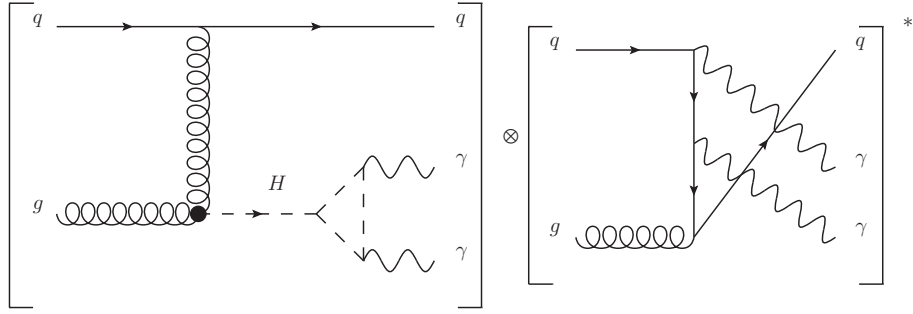}
\hspace{0.5cm} 
\caption{\label{fig:qgall} \small{Sample of Feynman diagrams contributing to the interference}}
\end {center}
\end{figure}

The amplitudes of the $qg$ and $q\bar{q}$ initiated contributions to the interference were calculated using the Mathematica package FeynArts \cite{Hahn:2000kx}, and the analytical manipulation to obtain the final squared matrix element of the complete interference was done with the help of the package FeynCalc  \cite{Mertig:1990an}. A sample of the Feynman diagrams for the $qg$ interference channel are shown in Figure \ref{fig:qgall}. The diagrams and amplitudes of the remaining channels can be obtained by performing the corresponding crossings. 

It is worth noticing that, compared to the $gg$ subprocess, there is an additional parton in the final state in the new channels. This parton  has to be integrated out to evaluate the impact on the cross section, and its appearance might provide the wrong impression that the contribution is next-to-leading order-like. However, since signal and background amplitudes develop infrared singularities in different kinematical configurations, the interference is finite after phase space integration and behaves as a true tree-level contribution, with exactly the same power of the coupling constant as the one arising from the gluon-gluon interference channel.

For a phenomenological analysis of the results we need to perform a convolution of the partonic cross-section with the parton density functions. We use the MSTW2008 LO set \cite{Martin:2009iq}  (five massless flavours are considered), and the one-loop expression of the strong coupling constant, setting the default factorization and renormalization scales equal to the diphoton invariant mass ($\mu_F=\mu_R= M_{\gamma\gamma}$). For the sake of simplicity, the production amplitudes are computed within the effective Lagrangian approach for the $ggH$ coupling (relying on the infinite top mass limit), approximation known to work at the few percent level for the process of interest. The decay into two photons is treated exactly (using $m_t=172.5\,\text{GeV}$, $m_b=4.75\,\text{GeV}$, $m_c=1.40\,\text{GeV}$ \cite{Dittmaier:2011ti,Dittmaier:2012vm}, $m_\tau=1.776\,\text{GeV}$, $m_W=80.395\,\text{GeV}$ \cite{PDG}) and we set $\alpha=1/137$.
For the continuum background $gg\to \gamma \gamma $ we consider five massless flavours.
For the Higgs boson we use $m_H=125\,\text{GeV}$ and $\Gamma_H=4.2\,\text{MeV}$. For all the histograms we present in this section, an asymmetric cut is applied to the transverse momentum of the photons: $p_{T, \gamma}^{hard (soft)} \geq 40(30) \, \text{GeV}$. Their pseudorapidity is constrained to $|\eta_\gamma | \leq 2.5$.
We also implement the standard isolation prescription for the photons,  
requesting that the transverse hadronic energy deposited within a cone of size
$R=\sqrt{\Delta \phi^2+\Delta \eta^2}<0.4$ around the photon should satisfy $
p_{T,had} \leq 3\,\text{GeV}$. Furthermore, we reject all the events with $R_{\gamma\gamma} < 0.4$. 
The effect of the precise definition of the isolation prescription is negligible since no final state photon-quark singularities appear at the level of the interference. 
Therefore, almost the same results are obtained if the smooth isolation prescription  \cite{Frixione:1998jh} is implemented. 

\begin{figure} 
\begin {center}
\includegraphics[scale=1.4]{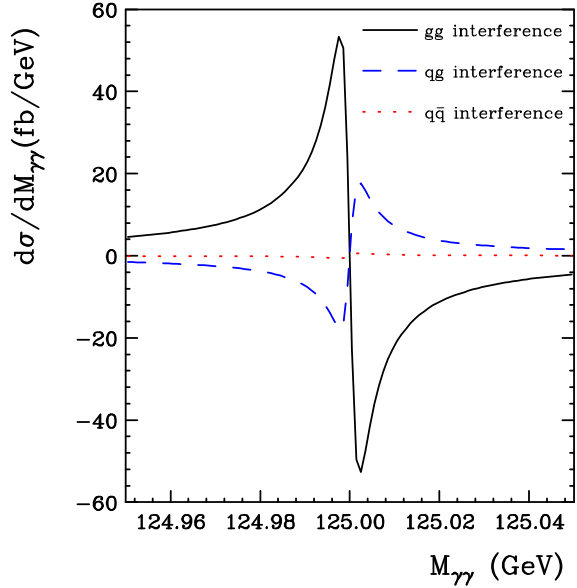}
\hspace{0.5cm} 
\caption{\label{fig:int_singauss} \small{Diphoton invariant mass distribution for the interference terms. The solid line is the $gg$ channel contribution, the dotted one the $qg$ channel, and dashed the $q \bar q$.}}
\end {center}
\end{figure}

In Figure \ref{fig:int_singauss} we show the three contributions to the full signal-background interference as a function of the diphoton invariant mass $M_{\gamma\gamma}$ after having implemented all the cuts mentioned above. The $gg$ term (solid line) represents the dominant channel, while the $qg$ contribution (dashed) is about 3 times smaller in absolute magnitude, but as we can observe, has the same shape but opposite sign to the $gg$ channel. The $q\bar{q}$ contribution (dotted) is two orders of magnitude smaller than the $gg$ one, and with the same sign of $qg$. The position of the maximum and minimum of the distribution are located near $M_{\gamma\gamma}=M_H \pm \Gamma_H/2$, generating a shift which at this level remains at ${\cal O}(1\, {\rm MeV})$.
\begin{figure} 
\begin {center}
\includegraphics[scale=1.4]{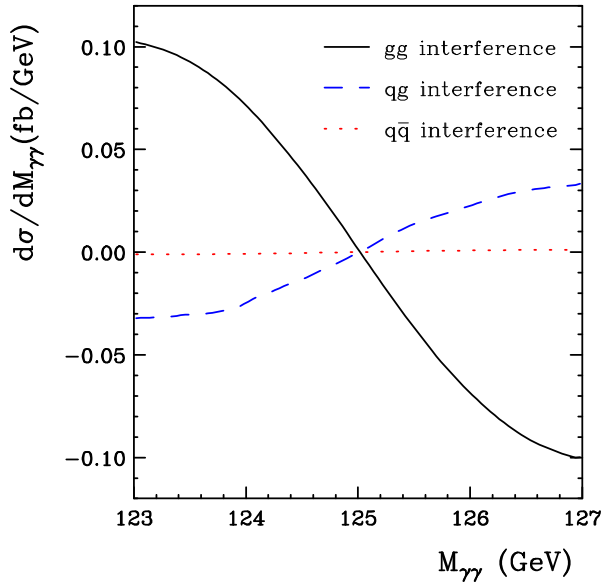}
\hspace{0.5cm} 
\caption{\label{fig:int_gauss} \small{Diphoton invariant mass distribution for the interference terms including the smearing effects which simulate the detector (gaussian function of width $1.7 \, \text{GeV}$). The solid line represents the $gg$ channel contribution, the dashed line represents the $qg$ channel, and the dotted one, the $q \bar q$.}}
\end {center}
\end{figure}

To simulate the smearing effects introduced by the detector, we convolute the cross-section with a gaussian function of mass resolution width $\sigma_{\text{MR}}=1.7\,\text{GeV}$ following the procedure of Ref.\cite{Martin:2012xc}. The corresponding results are presented in Figure \ref{fig:int_gauss}, where we can observe that the magnitude of the interference is reduced, and the position of the peak (and dip) is moved as much as $2 \, \text{GeV}$. The main reason for this shift is the highly antisymmetric nature of the interference, which is enhanced by the convolution with the gaussian function. The precise value of the displacement is proportional to both the width of the gaussian, and the absolute magnitude and sign of the interference, resulting in a shift towards larger invariant masses for the $qg$ and $q \bar q$ channels.

\begin{figure} 
\begin {center}
\includegraphics[scale=1.4]{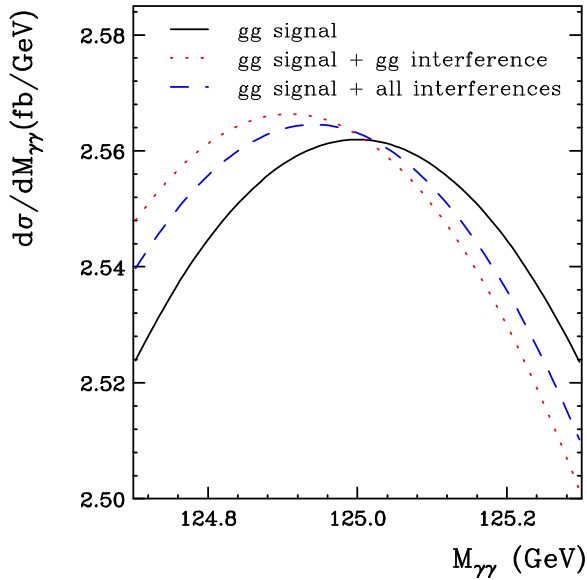}
\hspace{0.5cm} 
\caption{\label{fig:shift} \small{Diphoton invariant mass distribution including the smearing effects of the detector (gaussian function of width $1.7 \, \text{GeV}$). The solid line corresponds to the signal-only contribution. The dotted line corresponds to the distribution  after adding the $gg$ interference term, and the dashed line represents the complete Higgs signal plus all three interference contributions ($gg$, $qg$ and $q \bar q$).}}
\end {center}
\vspace{-1cm}
\end{figure}

In order to quantify the physical effect of the interferences in the diphoton invariant mass spectrum, we present in Figure \ref{fig:shift} the corresponding results after adding the Higgs signal. The solid curve corresponds to the signal cross-section, without the interference terms, but including the detector smearing effects. As expected, the (signal) Higgs peak remains at $M_{\gamma\gamma}=125 \, \text{GeV}$. When adding the $gg$ interference term, we observe a shift on the position of the peak around $90 \, \text{MeV}$ towards a lower mass (dotted), compatible with Ref.\cite{Martin:2012xc}. If we also add the $qg$ and $q \bar q$ contributions (dashed), the peak is shifted around $30 \, \text{MeV}$ back towards a higher mass region because of the opposite sign of the amplitudes. 
Given the fact that $q\bar{q}$ and $qg$ channels involve one extra particle in the final state, one might expect their contribution to be even more relevant for the corresponding interference in the process $pp \rightarrow H(\rightarrow \gamma\gamma) + \text{jet}$, since the {\it usually dominant} $gg$ channel starts to contribute at the next order in the strong coupling constant for this observable.

\begin{figure} 
\begin {center}
\includegraphics[scale=1]{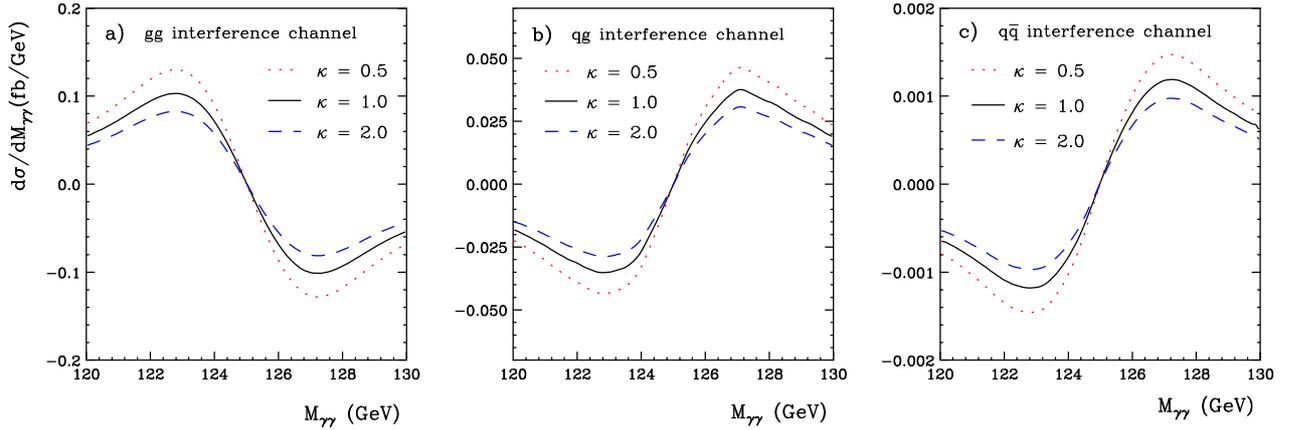}
\hspace{-0.3cm} 
\caption{\label{fig:Scale_Dependence} \small{Scale dependence of the diphoton invariant mass distribution for interference terms. The $gg$ channel contribution is shown in (a), $qg$ in (b) and $q \bar q$ in (c). Factorization and renormalization scales are varied simultaneously as $\mu_{F}=\mu_{R}= \kappa \,M_{\gamma\gamma}$ in the range $\frac{1}{2}\,\leq \kappa \leq 2$.}}
\end {center}
\end{figure}

It is worth noticing that the results presented here are plain LO in QCD. Given the fact that very large K-factors are observed in both the signal and the background, one might expect a considerable increase in the interference as well. 
As a first attempt to try to quantify the theoretical uncertainties, we study the   factorization and renormalization scale dependence of the interferences by varying both scales (simultaneously) from $\frac{1}{2} M_{\gamma\gamma}$ to $2 M_{\gamma\gamma}$. The results are shown in Figure \ref{fig:Scale_Dependence}, were we present the contribution from each channel including the smearing effect with $\sigma_{\text{MR}}=1.7\,\text{GeV}$.  Very large variations are observed, reaching up to $45 \%$ at the peaks in all channels.

Unfortunately, it is known that the LO scale variation for both signal and background in Higgs production is not enough to estimate the size of the missing higher order corrections. Therefore one might consider the result of this variation only as a lower bound on the contribution of higher order terms for the interference.

While calculating the interference to the same accuracy of the signal and background seems hard to be achieved at the present time, a prescription to estimate the uncertainty on the evaluation of the interference and a way to combine it with more precise higher order computations for signal and background for $gg\rightarrow ZZ$ was recently presented in \cite{Passarino:2012ri}. The procedure can be easily extended to the case presented here by including all possible initial state channels.

Finally, we would like to emphasize that a more realistic simulation of the detector effects should be performed in order to obtain reliable predictions and allow for a direct comparison with the experimental data.
The Fortran code to compute the interferences presented in this paper can be obtained upon request from the authors.

{\bf Acknowledgements}. We thank L. Cieri and the participants of the Second  Phenomenology Workshop (Tandil 2012) 
for discussions and contributions at the various stages of this work, and
L. Dixon and S. Martin for helpful comments and comparisons.
This work was supported in part by UBACYT, CONICET, ANPCyT and the Research Executive Agency (REA) of the European Union under the Grant Agreement number PITN-GA-2010-264564 (LHCPhenoNet).


\begin{thebibliography}{90}


\bibitem{Martin:2012xc}
  S.~P.~Martin,
  Phys.\ Rev.\ D {\bf 86} (2012) 073016
  [arXiv:1208.1533 [hep-ph]].


\bibitem{:2012gu}
  S.~Chatrchyan {\it et al.}  [CMS Collaboration],
  Phys.\ Lett.\ B {\bf 716} (2012) 30
  [arXiv:1207.7235 [hep-ex]].

\bibitem{:2012gk} 
  G.~Aad {\it et al.}  [ATLAS Collaboration],
  Phys.\ Lett.\ B {\bf 716}, 1 (2012)
  [arXiv:1207.7214 [hep-ex]].
  


\bibitem{ATLASTDR}
``ATLAS: Detector and physics performance technical design report. Volume 2,''
  CERN-LHCC-99-15.


\bibitem{Georgi:1977gs}
  H.~M.~Georgi, S.~L.~Glashow, M.~E.~Machacek and D.~V.~Nanopoulos,
  Phys.\ Rev.\ Lett.\  {\bf 40} (1978) 692.

\bibitem{Dawson:1990zj}
  S.~Dawson,
  Nucl.\ Phys.\  B {\bf 359} (1991) 283.

\bibitem{Djouadi:1991tk}
A.~Djouadi, M.~Spira and P.~M.~Zerwas,
Phys.\ Lett.\ B {\bf 264} (1991) 440.



\bibitem{Spira:1995rr}
  M.~Spira, A.~Djouadi, D.~Graudenz and P.~M.~Zerwas,
  Nucl.\ Phys.\  B {\bf 453} (1995) 17.


\bibitem{Harlander:2002wh}
  R.~V.~Harlander and W.~B.~Kilgore,
  Phys.\ Rev.\ Lett.\  {\bf 88} (2002) 201801.

\bibitem{Anastasiou:2002yz}
  C.~Anastasiou and K.~Melnikov,
  Nucl.\ Phys.\  B {\bf 646} (2002) 220.

\bibitem{Ravindran:2003um}
  V.~Ravindran, J.~Smith and W.~L.~van Neerven,
  Nucl.\ Phys.\  B {\bf 665} (2003) 325.



\bibitem{Anastasiou:2005qj}
  C.~Anastasiou, K.~Melnikov and F.~Petriello,
  Phys.\ Rev.\ Lett.\  {\bf 93} (2004) 262002,
  Nucl.\ Phys.\  B {\bf 724} (2005) 197.

\bibitem{Anastasiou:2007mz}
  C.~Anastasiou, G.~Dissertori and F.~Stockli,
  JHEP {\bf 0709} (2007) 018.


\bibitem{Catani:2007vq}
  S.~Catani and M.~Grazzini,
  Phys.\ Rev.\ Lett.\  {\bf 98} (2007) 222002.

\bibitem{Grazzini:2008tf}
  M.~Grazzini,
  JHEP {\bf 0802} (2008) 043.



\bibitem{deFlorian:2012yg}
  D.~de Florian and M.~Grazzini,
  Phys.\ Lett.\ B {\bf 718} (2012) 117
  [arXiv:1206.4133 [hep-ph]].




\bibitem{ew}
U.~Aglietti, R.~Bonciani, G.~Degrassi and A.~Vicini,
Phys.\ Lett.\ Â B {\bf 595} (2004) 432;
G.~Degrassi and F.~Maltoni,
Phys.\ Lett.\ Â B {\bf 600} (2004) 255;
U.~Aglietti, R.~Bonciani, G.~Degrassi and A.~Vicini,
arXiv:hep-ph/0610033.


\bibitem{Actis:2008ug}
S.~Actis, G.~Passarino, C.~Sturm and S.~Uccirati,
Phys.\ Lett.\  B {\bf 670} (2008) 12;
 Nucl.\ Phys.\  B {\bf 811} (2009) 182.



\bibitem{Catani:2003zt}
S.~Catani, D.~de Florian, M.~Grazzini and P.~Nason,
JHEP {\bf 0307} (2003) 028.

\bibitem{Dittmaier:2011ti} 
  S.~Dittmaier {\it et al.}  [LHC Higgs Cross Section Working Group Collaboration],
  arXiv:1101.0593 [hep-ph].

\bibitem{Dittmaier:2012vm} 
  S.~Dittmaier, {\it et al.} [LHC Higgs Cross Section Working Group Collaboration],
  arXiv:1201.3084 [hep-ph].


\bibitem{Ellis:1975ap}
  J.~R.~Ellis, M.~K.~Gaillard and D.~V.~Nanopoulos,
  Nucl.\ Phys.\ B {\bf 106}, 292 (1976).

\bibitem{Shifman:1979eb}
  M.~A.~Shifman, A.~I.~Vainshtein, M.~B.~Voloshin and V.~I.~Zakharov,
  Sov.\ J.\ Nucl.\ Phys.\  {\bf 30}, 711 (1979)
  [Yad.\ Fiz.\  {\bf 30}, 1368 (1979)].






\bibitem{Zheng:1990qa}
H.-Q. Zheng and D.-D. Wu, 
  {Phys. Rev. {\bf D42} (1990)
   3760--3763}.

\bibitem{Djouadi:1990aj}
A.~Djouadi, M.~Spira, J.~van~der Bij, and P.~Zerwas, 
   {Phys. Lett. {\bf B257}
  (1991)  187--190}.

\bibitem{Dawson:1992cy}
S.~Dawson and R.~Kauffman, 
  {Phys. Rev. {\bf D47} (1993)
   1264--1267}.

\bibitem{Djouadi:1993ji}
A.~Djouadi, M.~Spira, and P.~Zerwas, 
  {Phys. Lett. {\bf B311} (1993)  255--260},
  arXiv:hep-ph/9305335[hep-ph]. 

\bibitem{Melnikov:1993tj}
K.~Melnikov and O.~I. Yakovlev, 
 {Phys. Lett. {\bf B312}
  (1993)  179--183}, 
  arXiv:hep-ph/9302281 [hep-ph].

\bibitem{Inoue:1994jq}
M.~Inoue, R.~Najima, T.~Oka, and J.~Saito, 
  {Mod. Phys. Lett. {\bf A9}
  (1994)  1189--1194}.




\bibitem{Catani:2011qz}
  S.~Catani, L.~Cieri, D.~de Florian, G.~Ferrera and M.~Grazzini,
  Phys.\ Rev.\ Lett.\  {\bf 108} (2012) 072001
  [arXiv:1110.2375 [hep-ph]].
  
  
\bibitem{Goria:2011wa}
  S.~Goria, G.~Passarino and D.~Rosco,
  Nucl.\ Phys.\ B {\bf 864} (2012) 530
  [arXiv:1112.5517 [hep-ph]].


\bibitem{Dixon:2003yb}
  L.~J.~Dixon and M.~S.~Siu,
  Phys.\ Rev.\ Lett.\  {\bf 90} (2003) 252001
  [hep-ph/0302233].

\bibitem{Dicus:1987fk}
  D.~A.~Dicus and S.~S.~D.~Willenbrock,
  Phys.\ Rev.\ D {\bf 37} (1988) 1801.


\bibitem{Hahn:2000kx}
  T.~Hahn,
  Comput.\ Phys.\ Commun.\  {\bf 140} (2001) 418
  [hep-ph/0012260].

\bibitem{Mertig:1990an}
  R.~Mertig, M.~Bohm and A.~Denner,
  Comput.\ Phys.\ Commun.\  {\bf 64} (1991) 345.




\bibitem{Martin:2009iq}
  A.~D.~Martin, W.~J.~Stirling, R.~S.~Thorne and G.~Watt,
  Eur.\ Phys.\ J.\ C {\bf 63} (2009) 189
  [arXiv:0901.0002 [hep-ph]].


\bibitem{PDG}
J. Beringer et al. (Particle Data Group), Phys. Rev. D86, 010001 (2012).


\bibitem{Frixione:1998jh}
  S.~Frixione,
  Phys.\ Lett.\ B {\bf 429} (1998) 369
  [hep-ph/9801442].


  
\bibitem{Passarino:2012ri}
  G.~Passarino,
  JHEP {\bf 1208} (2012) 146
  [arXiv:1206.3824 [hep-ph]].
  
\end{thebibliography}
\end{document}